\documentclass[sigconf,nonacm]{acmart}

\AtBeginDocument{%
  }

\setcopyright{acmlicensed}
\copyrightyear{2024}
\acmYear{2024}
\acmDOI{XXXXXXX.XXXXXXX}

\acmConference[SE 2030]{International Workshop on Software Engineering in 2030}{November 2024}{Puerto Galinàs (Brazil)}
\acmISBN{}

\begin{document}

\title{Model-based Maintenance and Evolution with GenAI: A Look into the Future}

\author{Luciano Marchezan} 
\orcid{0000-0003-3096-580X}
\affiliation{%
  \institution{ISSE - Johannes Kepler University}
    \streetaddress{Altenberger Straße 69}
  \city{Linz} 
  \country{Austria}
  \postcode{4040}
}

\author{Wesley K. G. Assunção}
\orcid{0000-0002-7557-9091}
\affiliation{%
  \institution{North Carolina State University}
  \city{Raleigh} 
  \country{USA}
}


\author{Edvin Herac}
\orcid{0000-0002-9638-8569}
\author{Alexander Egyed}
\orcid{0000-0003-3128-5427}
\affiliation{%
  \institution{ISSE - Johannes Kepler University}
  \streetaddress{Altenberger Straße 69}
  \city{Linz} 
  \country{Austria}
  \postcode{4040}
}

\renewcommand{\shortauthors}{Marchezan et al.}

\begin{abstract}
Model-Based Engineering (MBE) has streamlined software development by focusing on abstraction and automation. The adoption of MBE in Maintenance and Evolution (MBM\&E), however, is still limited due to poor tool support and a lack of perceived benefits. We argue that Generative Artificial Intelligence (GenAI) can be used as a means to address the limitations of MBM\&E. In this sense, we argue that GenAI, driven by Foundation Models, offers promising potential for enhancing MBM\&E tasks. With this possibility in mind, we introduce a research vision that contains a classification scheme for GenAI approaches in MBM\&E considering two main aspects: (i) the level of augmentation provided by GenAI and (ii) the experience of the engineers involved. We propose that GenAI can be used in MBM\&E for: reducing engineers' learning curve, maximizing efficiency with recommendations, or serving as a reasoning tool to understand domain problems. Furthermore, we outline challenges in this field as a research agenda to drive scientific and practical future solutions. With this proposed vision, we aim to bridge the gap between GenAI and MBM\&E, presenting a structured and sophisticated way for advancing MBM\&E practices.
\end{abstract}


\maketitle

\section{Introduction}

Model-based engineering (MBE) and its sub-fields such as model-driven software engineering (MDSE)~\cite{Brambilla2017} have shown many benefits over the years, allowing engineers to tame complexity using abstraction (domain-modeling), automation (model transformation), and simulation (validation), which reduces the effort to develop software~\cite{Wasowski2023}. Instead of focusing on technological aspects, by using MBE, engineers can focus on domain concerns~\cite{Fowler2010}. 
In this context, MBE can be used in different software engineering activities, such as maintenance and evolution~\cite{Torres2020,Raibulet2017}. Considering the maintenance part, MBE approaches support consistency checking~\cite{Marchezan2023SEIP}, repairing~\cite{Marchezan2022}, and overall quality assurance~\cite{Torres2020}. For the evolution part, approaches target co-evolution~\cite{Homolka2024}, software modernization~\cite{Raibulet2017}, reengineering~\cite{Acher2023a}, among others. MBE for maintenance and evolution (MBM\&E), however, has received less attention from both research and practice in comparison to programming or testing~\cite{Kosar2016}.

This is a missed opportunity as the benefits of MBM\&E had shown promising results~\cite{Torres2020,Raibulet2017}. The reasons for the lack of adoption of MBM\&E in practice include the lack of proper tools and the clear understanding of the benefits that companies can gain~\cite{Ozkaya2021}. For example, one of the most prominent aspects of MBE is to reduce the time of tedious activities such as refactoring~\cite{Bettini2022}. Still, there are different challenges faced when dealing with MBM\&E, such as integrating non-intrusive approaches into a well-established software development process, which requires high generalization while allowing flexibility and customization. Furthermore, introducing newcomers to MBM\&E, especially to MDSE, is not trivial as the learning curve of such approaches can be steep~\cite{Bucchiarone2020}. 

This low adoption of MBM\&E in practice raises the question: \textit{What is required to make MBM\&E well-adopted in the industry?} We argue that there is potential to explore the recent advances in the Generative Artificial Intelligence (GenAI) field~\cite{Ebert2023} to address this question. These advances introduced several opportunities for software engineering~\cite{Greiner2024,Ebert2023}, especially with the possibilities created by Foundation Models (FM)~\cite{Wei2022}. Their usefulness is evident in different software activities disciplines~\cite{Fan2023}, such as requirements~\cite{Preda2024}, implementation~\cite{Rajbhoj2024,Tufano2023}, and education~\cite{Haindl2024}. 
While beneficial, the use of GenAI in software engineering has focused on source code or requirements. 
Despite this, there is potential to use GenAI for MBM\&E as a means to overcome the challenges and low adoption of the field~\cite{Bucchiarone2020}. This is evidenced by initial efforts from researchers to adopt GenAI techniques to MBM\&E~\cite{Acher2023a,Liu2024,Cabot2022}.

In this paper, we propose a research vision for the next years considering the potential of applying GenAI for MBM\&E. Our research vision has two contributions. First, we propose an initial classification scheme that considers the level of augmentation provided by GenAI approaches in contract with the experience possessed by engineers, as a way of guiding the research on the field to focus on four different possibilities: assisting, leveling up, reasoning, and automating (Section~\ref{sec:classification}). 
Second, we suggest a possible research agenda to spark new work in the field of MBM\&E + GenAI. In this agenda, we discuss the possible challenges and open opportunities that can be explored in the field in the next years (Section~\ref{sec:agenda}). 


\section{Background}



\noindent \textbf{Model-based Maintenance and Evolution.}
In this paper, we focus on MBE as in ``approaches that use models'' instead of MDSE where ``the model is the key artifact''. The ideas presented in this paper, however, can also be adapted for MDSE. In this sense, MBE approaches and tools are prominent in many engineering fields~\cite{Hutchinson2011,Iung2020} with maintenance and evolution being an important aspect of these approaches~\cite{Torres2020,Raibulet2017}. The main reason for this is the abstraction of concepts provided by models, which allows domain experts to engineer solutions without necessarily understanding other activities such as programming and testing~\cite{Wegeler2013}. An industry-proven example of this possibility are domain-specific languages (DSL)~\cite{Fowler2010}, which are widely used in many domains~\cite{Wasowski2023}. 

When considering maintenance and evolution, MBE or MDSE solutions such as DSLs, have been proposed over the years~\cite{Torres2020}. When compared to other software engineering disciplines, e.g., design or testing, model-based solutions for maintenance and evolution are less adopted in practice~\cite{Kosar2016}. This is a contradictory situation since maintenance and evolution are acknowledged to be highly complex activities that can require up to 80\% of the software costs~\cite{Pigoski1996, Canfora2011} with more model-based support for them being desired by engineers~\cite{David2023}.
Reasons for this low adoption include the lack of tool support that provides clear benefits to motivate companies to invest time and money~\cite{Ozkaya2021}. These benefits include improving the quality of results, helping engineers with tedious or repetitive tasks, aiding the thinking process, or even reducing the learning curve required to adopt model-based solutions~\cite{Bucchiarone2020}. We argue that these benefits are hard to achieve for any proposed model-based solution. When considering model-based maintenance and evolution (MBM\&E), these are even more challenging as maintenance and evolution are overlooked phases when compared to requirements engineering or implementation~\cite{Jongeling2019,Assunccao2017}. Although maintenance and evolution can be investigated separately, in this paper we look into them as one aspect of the software engineering life cycle. Thus, we talk of \textit{maintenance as a means for evolution}. Hence, we focus on the MBM\&E aspect of software engineering. Thus, we argue that there is potential to be explored by investigating how GenAI~\cite{Ebert2023} can be used to address  MBM\&E limitations. 

\vspace{2mm}
\noindent \textbf{GenAI in Software Engineering.}
GenAI, more specifically, Foundation models (FM)~\cite{Wei2022} have shown their benefits in addressing different software engineering tasks~\cite{Fan2023}, such as source code related tasks~\cite{Rajbhoj2024,Tufano2023}, requirements engineering~\cite{Preda2024}, and education~\cite{Haindl2024}. In this sense, FMs have been used not only in scientific but also in practical contributions, such as GitHub copilot\footnote{GitHub copilot: \url{https://github.com/features/copilot}} and Code4Me\footnote{Code4M: \url{https://code4me.me/}} for code specific tasks, or GitHub workspaces\footnote{GitHub workspaces: \url{https://githubnext.com/projects/copilot-workspace/}} for reasoning about the whole development life-cycle.
Work considering FMs for MBE has shown how MBE can be used to improve aspects of FMs~\cite{Naveed2024}, e.g., prompt engineering~\cite{Clariso2023}. We argue, however, that the real benefit for MBE is leveraging FMs to be helpful for model-based related techniques. The use of AI for model-based techniques is not novel, been explored in the past~\cite{Kessentini2018,Kessentini2019,Desai2016,Virmajoki2022}. The use of GenAI for MBE, however, is still in its early stages, with promising results~\cite{Acher2023a,Acher2023b,Liu2024,Netz2024,Fill2023,Camara2023,Galindo2023}. That is the reason why we envision that in the next years, the application of GenAI for MBE will only grow, which leads to the possibilities of addressing the challenges of MBM\&E. In the next section, we describe different aspects of MBM\&E that we believe can be explored with the use of GenAI.

\section{Generative AI for MBM\&E}
\label{sec:classification}
\begin{figure}
    \centering
    \includegraphics[width=.6\columnwidth]{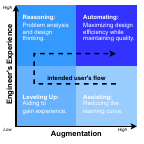}
\caption{Classification scheme of GenAI applied to MBM\&E}
    \label{fig:scheme}
\end{figure}

Similar to how GenAI is used for different software engineering activities~\cite{Fan2023}, its use for MBM\&E can vary. We envision two main aspects that may be prominent and important for the scientific and practical communities: the level of augmentation, and the level of engineers' experience (their relationships are illustrated in Figure~\ref{fig:scheme}). First, the augmentation level describes how much the use of GenAI replaces the manual work performed by engineers with a high augmentation nearing the border of automation (completely replacing humans). We believe that in the next five to ten years, GenAI will not reach a point where humans can be replaced in any given MBM\&E task (reasons are discussed in Section~\ref{sec:agenda}). Thus, even with a high level of augmentation, we still envision humans in the loop for all MBM\&E tasks. Moreover, we envision that proposals on this area should focus on the so-called Software 4.0 \textit{Agentware} layer (i.e., interacting directly with users while reducing the complexity of the back-end), with GenAI models being the mechanism to drive the software agents involved~\cite{Hassan2024}. Second, the level of engineers' experience is related to the expertise that an engineer has in the domain and with the tools needed to perform MBM\&E tasks (both GenAI or non-GenAI tools). While we focus on MBM\&E aspects, we argue that our classification may also be considered important for GenAI approaches as a whole. When compared to programming, for example, adopting MBM\&E in practice concerns different challenges, e.g., a steeper learning curve as programming is more taught and adopted in practice. 

Figure~\ref{fig:scheme} illustrates our classification scheme, based on the combinations of high/low augmentation and experience that are used to define four distinct research directions considering GenAI for MBM\&E: \textit{Assisting}, \textit{Leveling Up}, \textit{Reasoning}, and \textit{Automating}. We believe that this order is the intended flow for engineers as newcomers can relate to high augmentation during the learning phase moving to low augmentation to improve their MBE-based skills, such as thinking about design problems. This will lead them to increase their experience in the domain/tools, moving to the reasoning part where GenAI can aid with problem investigation and understanding (something expected from senior engineers). The last step would be increasing the efficiency of experienced engineers with high augmentation, i.e., automation. Details are given in what follows.

\subsection{Assisting}
\textbf{Motivation}: MBM\&E techniques and technologies can be overwhelming for newcomers, especially considering MDSE~\cite{Brambilla2017}, as the introduction of additional artifacts (different models) and technologies can be a burden for engineers. In this sense, GenAI can be helpful to reduce the learning curve working as an assistant, such as a chatbot integrated into a tool. 
\textbf{Focus}: This direction aims at the assistance to facilitate newcomers to gain rapid knowledge of the domain/tools. Moreover, approaches can focus on assisting with more general tasks, such as describing how reengineering or co-evolution steps should be planned and executed, possibly even executing them to show possible output to the engineers. 
\textbf{Augmentation}: High augmentation is recommended for newcomers. \textbf{Initial effort}: This topic of GenAI is overlooked considering MBM\&E. Even for software engineering as a whole, this topic had just started to show initial results~\cite{Dominic2020,Nam2024}. Thus, there are plenty of opportunities for research in this area, despite the challenges imposed (as discussed in Section~\ref{sec:agenda}).

\subsection{Leveling Up}
\textbf{Motivation}: Once newcomers are used to the MBM\&E techniques, they start getting experienced in their tasks. Here, GenAI can be beneficial to improve the performance, while reducing the augmentation to require engineers to improve their skills (especially the design skills).
\textbf{Focus}: Aims at aiding the engineer to increase their experience in the domain and tools related to their tasks. Instead of working as an assistant, approaches should focus on maximizing the thinking process of engineers with guided help regarding specific tasks only (e.g., automating model co-evolution by recommending possibilities instead of just executing them).
\textbf{Augmentation}: Low augmentation is desired at this point, to allow engineers to increase their ``hands-on'' approach, which leads to gaining more knowledge and experience on the domain or the tool related to their MBM\&E role/task.
\textbf{Initial effort}: Similar to \textit{Assisting}, this category has been overlooked by the MBE community. Approaches that help with model auto-completion are one example of possible opportunities~\cite{Liu2024} as engineers are partially assisted during modeling or evolution steps.

\subsection{Reasoning}\label{sec:reasoning}
\textbf{Motivation}: While GitHub co-pilot has been used as a tool to improve the efficiency of programming tasks, one of its main benefits is helping programmers get unstuck when they do not know how to progress (similar to using pair programming~\cite{Imai2022}). This benefit not only improves efficiency, but it also aids programmers with their thinking process, especially because for most of the generated code, they are required to do refactoring~\cite{Arghavan2023}. In the same sense, we argue that solutions for MBM\&E must be designed and implemented to aid engineers during the reasoning process. This is especially important for more experienced engineers who are in charge of model-based tasks involving a higher amount of thinking process. 
\textbf{Focus}: We envision that GenAI can aid engineers with the reasoning by providing suggestions regarding the understandability of the domain problems (e.g., recommending similar problems and solutions) related to different MBM\&E tasks, e.g., model refactoring due to inconsistencies, or model evolution to address legacy problems. 
\textbf{Augmentation}: Since the focus is on problem reasoning, the augmentation should be low, as generation or automatic refactoring may bias engineers during the thinking process. 
\textbf{Initial effort}: Work has been seen with regard to investigating reengineering problems~\cite{Acher2023a,Acher2023b} or generation of domain representations~\cite{Fill2023}.

\subsection{Automating}
\textbf{Motivation}: One of the major benefits of using GenAI is to reduce the effort of tasks, aiding engineers to be more efficient. Similarly, the quality of the produced artifacts can be improved as engineers can focus on quality constraints rather than on the actual creation of the artifacts since they can be generated by AI. While we believe that most tasks involving GenAI must keep the human in the loop, we also see the potential for increased augmentation in specific (less critical) tasks of MBM\&E. One example is transformation tasks, such as transforming a textual description of a model inconsistency~\cite{Marchezan2023SEIP} into a visual representation, highlighting the inconsistent parts of the model. While the overall task (repairing a model) is not augmented, the specific task (representing an inconsistency) is fully augmented by the use of GenAI.
\textbf{Focus}: We envision that automating specific MBM\&E tasks can reduce their effort and improve their quality. This may be achieved by, for example, integrating FM solutions in design or reengineering tools for suggesting model repair or refactoring, leading to new versions of models. 
\textbf{Augmentation}: The automation is partial to ensure that the human is still in the loop, something important due to the non-deterministic nature of GenAI. 
\textbf{Initial effort}: Examples of this direction have been proposed considering different topics such as generation of variability models for reengineering~\cite{Galindo2023, assunccao2017multi} or generation of complete new architectures for model-based evolution~\cite{Netz2024, Assuncao2020IST}. Similar to all other categories, there are still many opportunities open for research, as discussed in the next section.

\section{Research Agenda}\label{sec:agenda} 
\begin{figure}[tb]
    \centering
    \includegraphics[width=.95\columnwidth]{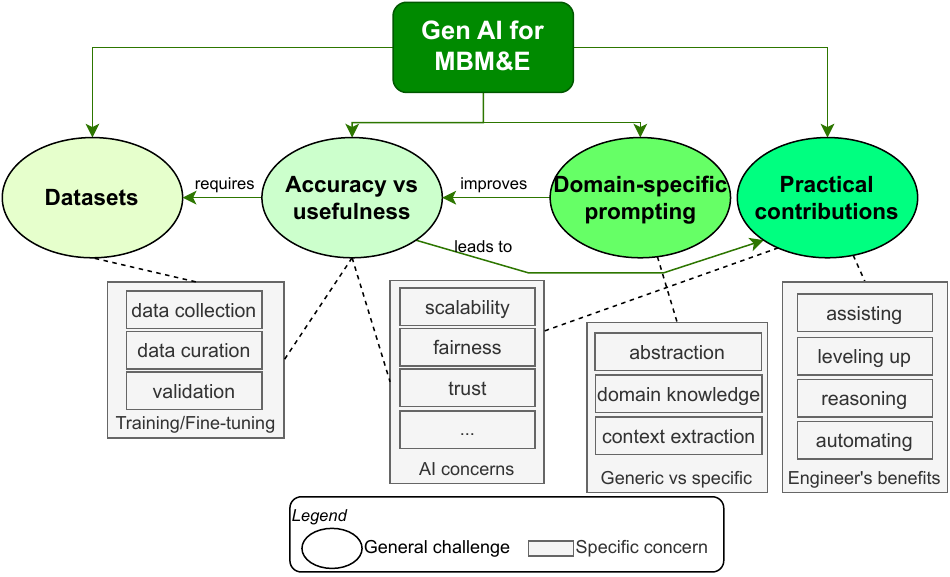}
    \caption{Conceptual map of the challenges related to GenAI for MBM\&E}
    \label{fig:map}
\end{figure}

The classification presented in Figure~\ref{fig:scheme} can be used as a starting point to systematically categorize current work and to plan future work in MBM\&E + GenAI. We have also to consider potential challenges that will be faced by researchers or practitioners going in this direction. Addressing these challenges can be seen as a possible research agenda for the next years in this field. The challenges presented in this section focus on the GenAI-related aspects described in Section~\ref{sec:classification}, while somehow intersecting the challenges of the MBE and MDSE communities as a whole~\cite{Bucchiarone2020}.
Figure~\ref{fig:map} illustrates the relationship between our proposed challenges and their specific concerns. Details about each challenge are provided next.

\subsection{The dataset problem for MBM\&E}
One issue that has always been a challenge for applying AI techniques in MBE is the lack of curated datasets for training and validation. Despite the effort of recent work~\cite{Lopez2022}, the MBE community still suffers from this limitation. When considering MBM\&E, this issue is even more challenging, as finding relevant maintenance and evolution data is more difficult even when considering source code. The generalizable nature of FMs reduces this challenge for MBM\&E, however, as shown in recent work on the topic~\cite{Camara2023} the quality of the generated outputs is still far from a satisfactory point. This means that \textit{although FMs can be used for MBM\&E, data is still needed for fine-tuning and improving the quality of the generated output.} Research has been initiated in this direction, for example in the consistency checking domain~\cite{Cabot2022,Abukhalaf2023}. Despite this, such a challenge opens different opportunities, summarized as a research question (RQ): How to collect and curate meaningful datasets for fine-tuning FMs for specific MBM\&E tasks such as maintenance, reengineering, evolution, and co-evolution of models?

\subsection{Accuracy vs usefulness}
The investigation of the dataset challenge is directly related to the discussion regarding the accuracy of the results in contrast with the usefulness of the approaches/tools. As stated before, despite GitHub copilot still producing borderline accuracy with its output~\cite{Nguyen2022}, its usefulness can be seen in both its popularity and its improvement towards aiding users to understand the code rather than just generating it~\cite{Arghavan2023}. We argue that the same applies to MBM\&E approaches based on GenAI, with a high emphasis on reasoning and comprehension (as discussed in Section~\ref{sec:reasoning}. Thus, research has to \textit{consider accuracy and usefulness together and find a balance between them}, focusing on the engineer's benefits. In this direction, we see the possible RQs: To what extent can GenAI models improve their accuracy for MBM\&E specific tasks? How can engineers most benefit from the use of GenAI in MBM\&E tasks?

\subsection{Domain-specific prompting}
The quality of the output does not rely solely on the dataset, but also on the quality of the given input, e.g., prompts in FMs. The more context (not only size but also reasoning) is given, the better the results produced by FMs~\cite{Berryman2025}. Thus, we see the creation of a sub-field focusing on what we call Domain-Specific Prompting (DSP). The goal is similar to how DSLs were introduced to reduce the effort of domain experts during modeling activities. Thus, \textit{DSP can be used for domain experts to focus on the key aspects of the domain needed for the prompting}, while not necessarily becoming prompt engineers. This direction has to consider different aspects of modeling such as the abstraction of concepts, capturing and communicating domain knowledge, and well as extracting the context of engineers to be used as part of the prompts. Initial effort has been made in this direction by integrating MBE into the prompt engineering to create DSLs for DSP~\cite{Clariso2023} or by using DSP for MBM\&E tasks~\cite{Abukhalaf2023}. This direction of research has a clear goal, which can be summarized by the following RQ: How can MBE techniques be used to systematically improve DSP?

\subsection{Practical contributions for MBM\&E}
Low-code tools are a good example of how MBE, MDSE in this case, can be used to automate software tasks. Similarly, low-modeling tools that use GenAI such as BESSER\footnote{BESSER: \url{https://modeling-languages.com/besser-released-open-source-low-code-tool/}} are gaining popularity due to their practical use. The low-modeling aspect is powered by GenAI models that assist users and automate their tasks (both categories belonging to the high augmentation aspect from Figure~\ref{fig:scheme}). Besides initial effort, different from low-coding, low-modeling tools powered by GenAI for practical use have just started to be developed and released. Furthermore, practitioners are still reluctant to use GenAI for MBE tasks~\cite{Ozkaya2021}. This shows that \textit{we are just paving the way towards the adoption in industry} that we aim for as part of this roadmap. Still, practical contributions focused on MBM\&E are lacking as the focus is on other disciplines such as requirements engineering or implementation. In this sense, we look into two possible RQs: To what extent can low-modeling MBM\&E tools powered by GenAI bring practical benefits for companies? How generalizable (or customizable) should GenAI solutions be for MBM\&E tasks?

Investigating these RQs would lead to additional challenges such as the scalability of the approaches, and AI-specific concerns such as fairness, and transparency, among others. However, we are still too far from consolidated solutions for MBM\&E to consider these concerns as major challenges. They do, however, need to always be considered when dealing with GenAI solutions. Despite this, the advance in GenAI and FMs may finally give researchers and practitioners the resources necessary to advance MBM\&E in a way that it gets the focus and attention that it needs, considering its crucial role in software and systems development. This will then lead to the possibility of supporting engineers throughout their intended flow of working with GenAI in MBM\&E, as presented in Figure~\ref{fig:scheme}.

\section{Conclusion}
The constant problem of overlooking MBM\&E tasks is damaging for both research and practice. These tasks are not trivial, since they require reasoning, time, effort, learning complex tools, and producing precise results. For these reasons, over the years they have received less attention in comparison to other software engineering disciplines such as requirements and implementation. We believe that the challenges related to the adoption of MBM\&E tasks can be addressed by using GenAI approaches in different ways. In this paper, we present a preliminary classification scheme to help describe the possibilities of these approaches considering the engineer's experience and the level of augmentation required. This leads to the four categories of Assisting, Leveling up, Reasoning, and Automating. Following this path, however, requires addressing different challenges related to the use of GenAI in an MBM\&E scenario, including lack of datasets, balancing accuracy with usefulness, the need and ways to do DSP, and how to achieve practical contributions that are interesting to the industry.

While this paper aims at being a research agenda regarding challenges and opportunities, we based our vision and insights on the literature. Thus, we argue that this paper is a first step towards a systematic review of the use and possibilities of GenAI applied to MBM\&E (or MBE as a whole). We plan to follow this path to consolidate and enhance our classification scheme, the challenges and opportunities described in this paper.

\begin{acks}
This research has been funded by the Austrian Science Fund (FWF, 10.55776/P34805) and by BMK, BMDW, and the State of Upper Austria in the frame of SCCH, part of the COMET Programme managed by FFG. 
\end{acks}

\bibliographystyle{ACM-Reference-Format}
\bibliography{references.bib}

\end{document}